\documentclass[10pt]{amsart}
\allowdisplaybreaks[1]

\usepackage{amssymb}
\usepackage{amsmath}
\usepackage{amscd}
\usepackage{graphicx}
\usepackage{epstopdf}

\theoremstyle{plain}
\newtheorem{theorem}{Theorem}

\theoremstyle{definition}

\newcommand{\ts}{\hspace{0.5pt}}

\newcommand{\RR}{\mathbb{R}\ts}

\newcommand{\ZZ}{\mathbb{Z}}

\newcommand{\TT}{\mathbb T}

\newcommand{\Rd}{{\mathbb R}^d}

\newcommand{\vL}{\varLambda}
\newcommand{\gL}{\Lambda}

\newcommand{\cL}{\mathcal{L}}

\newcommand{\CalS}{\mathcal{S}}

\newcommand{\card}{{\mathrm{card}}}

\newcommand{\Vol}{\ell}

\newcommand{\frq}{{\rm freq}}
\newcommand{\bc}{\boldsymbol{c}}
\newcommand{\bm}{\boldsymbol{m}}
\newcommand{\bw}{\boldsymbol{w}}
\newcommand{\bk}{\boldsymbol{k}}

\newcommand{\bz}{\boldsymbol{z}}
\newcommand{\ws}{\boldsymbol{ws}}

\begin{document}
\title[Weighted model sets and their higher point-correlations]
{Weighted model sets and their higher point-correlations}

\author{Xinghua Deng}
\address{Department of Mathematics and Statistics,
The Open University, Walton Hall,
Milton Keynes MK7 6AA,  United Kingdom}
\email{x.deng@open.ac.uk}

\author{Robert V.\ Moody}
\address{Department of Mathematics and Statistics,
University of Victoria,
PO Box 3060, STN CSC,
Victoria BC,ÊÊV8W 3R4,
Canada}
\email{rmoody@uvic.ca}

\date{\today}

\begin{abstract}
Examples of distinct weighted model sets with equal $2,3,4, 5$-point
correlations are given.
\end{abstract}

\maketitle

{\it AMS Classification Codes: 52C23, 51P05, 74E15, 60G55}

\section{Introduction}\label{intro}

This short paper can be thought of as an extension of \cite{DM2} where we have proved that a regular real model set with a real internal space is, up to translation and alterations of density zero, uniquely determined by its $2$- and $3$-point correlation measures. There we have also given an example in the case that the internal space is the product space of a real space and a finite group that shows that this result does not extend in general to more complicated internal spaces. More precisely, we have given an example of two distinct model sets
having equal $2$- and $3$-point correlation measures, created using a single cut and project scheme, that are not translationally equivalent, even allowing for alterations of density zero.

In this paper, extending the setting to {\em weighted} model sets, we offer examples of pairs of weighted model sets which have equal $2,3,4,5$-point correlation measures, created using a single cut and project scheme, that are not translationally equivalent. These pairs are created by imposing a $6$-colouring on a previously constructed aperiodic model set and then further imposing on them two different weighting schemes. The resulting weighted model sets are then different, not even having any weights
in common, but their correlations measures, up to the fifth one are identical. The relevance of this type of result to the theory of
long range-aperiodic order and crystallography and some comments on its history are discussed in \cite{DM2}.

The key to this result (and also to the previous examples for unweighted model sets) is to use corresponding results, developed in \cite{GM}, for one dimensional periodic
sets. There is a simple way in which to intertwine this periodicity with the structure of
an ordinary model set, and this leads to aperiodic weighted model sets with points of several types or colours which have the same coincidence of $2 ,3,4,5$-point correlations.

In brief outline, we start with a weighted periodic model set on the real line whose internal space is $\ZZ/N\ZZ$ for some $N >1$.
We let $\CalS=(\Rd\times H,\cL)$ be an arbitrary cut and project scheme, where $H$ is a locally compact Abelian group and $\cL$ is a lattice in $\Rd\times H$. As usual, we denote by $L$ the projection of $\cL$ into $\Rd$ and denote by $(\cdot)^\star$ the star map of $\CalS$ from $L$ to $H$. Now assume that there is a {\em surjective} homomorphism
$\alpha:L\longrightarrow \ZZ/N\ZZ$. Then there is a combined cut and project scheme specified by
$\CalS^e = (\Rd\times (H\times \ZZ/N\ZZ), \cL^e)$,where $\cL^e :=\{(x,(x^\star,\alpha(x))): x\in L\}$. We call $\alpha(x)$ the colour or type of $x\in L$.  We use this new cut and project scheme to create coloured model sets with several windows, each corresponding to a different colour. In order to get a weighted model set we weight the points of this coloured model set according to their type.

We assume that the reader is familiar with the basic theory of model sets \cite{MoodyNato}. In \S2 we briefly recall the definitions that we need and the important concepts of uniform distribution
and the point-correlation measures. Closely connected to correlation measures
are the pattern frequencies, which are generally more convenient for our purposes. In \S3 we consider discrete {\it periodic} point sets on the real line, regarding them as model sets, and present the general formula for their finite-point correlation measures. In \S4 we elaborate the method of building an {\it extented} weighted model set from a weighted periodic system and an aperiodic model set, and then determine the precise form of the resulting pattern frequencies. In \S5 we take an example from
\cite{GM} of two weighted periodic systems (based on $\ZZ/6\ZZ$) which have equal $2,3,4,5$-point correlation measures (we offer a short proof of this),  and then use it to show that our extension construction will produce any number of aperiodic weighted real model sets with equal $2,3,4,5$-point correlation measures. Finally,  by way of illustration, we offer an example of this using a $6$-colouring of the vertices of Franz G\"ahler's shield tiling.

\section{Model sets and correlations} \label{ms}

We work in $\RR^d$. The usual Lebesgue measure will be denoted by $\Vol$.
The open cube of side length $R$ centred at $0$ is denoted by $C_R$.
We begin with the  {\em cut and project scheme}
$\CalS = (\RR^d, H, \cL)$ consisting of a compactly generated locally compact Abelian group $H$ and a lattice $\cL \subset \RR^d \times H$ for which the
projection mappings
$\pi_1$ and $\pi_2$ from $\RR^d \times H$ onto
$\RR^d$ and $H$ are injective and have dense image respectively:

\begin{equation} \label{cpScheme}
   \begin{array}{ccccc}
      \RR^d & \stackrel{\pi_{1}}{\longleftarrow} &
        \RR^d \times H & \stackrel{\pi_{2}}
      {\longrightarrow} & H   \\
      &&  \cup \\
      L&\stackrel{\simeq}{\longleftrightarrow} & \cL &\qquad &\qquad \\
      x& \leftrightarrow & \tilde x & \mapsto & x^\star \quad.
   \end{array}
\end{equation}
Then $L:=\pi_1 (\cL)$ is isomorphic as a group to $\cL$ and we have the mapping
$(\cdot)^\star \! : \, L \longrightarrow H$, $x \mapsto \tilde x \mapsto x^\star$,
with dense image, defined by $\pi_2 \circ (\pi_1|_{\cL})^{-1}$.

The statement that $\cL$ is a lattice is equivalent to saying that it
is a discrete subgroup of $\RR^d \times H$ and that the
quotient group $\TT := (\RR^d \times H)/\cL$ is compact.
We let $\theta_H$ be a Haar measure on $H$, normalized so that
the product measure $\Vol \otimes \theta_H$ on $\RR^d \times H$ gives
total measure $1$ to a fundamental region of the lattice $\cL$. \footnote{Other normalizations are possible. The one we have chosen leads to the formula
of Thm.\ref{uniformDensity}. Other normalizations produce multiplicative
factors in this formula. None of this is relevant to what we need here.}
We let $\theta_\TT$
be the canonical Haar measure on $\TT$ whose total measure is $1$.

 For $W \subset H$,
\[ \vL(W):= \{ u\in L: u^\star \in W\} \,. \]
A set $W \subset H$ is called a {\em window} if
$\Sigma^\circ \subset W \subset \Sigma$ for some compact
set $\Sigma \subset H$ which satisfies $\overline{\Sigma^\circ} =\Sigma$.

{\em We shall assume throughout that all windows that we use have
boundaries of $\theta_H$-measure $0$}.

We will deal with multiple windows, and it is convenient to allow windows
to be empty (which is allowed by the definition).

A (regular) {\em model set} or {\em cut and project set} is a set of the form
$\gL= x+  \vL(W)$ where $W$ is a window and $x\in \RR^d$. In the sequel we
shall only have need of the simpler model sets of the form $\gL = \vL(W)$.
\smallskip

Model sets are uniformly distributed point sets:

\begin{theorem} \cite{Moody} \label{uniformDensity}
Let $W \subset H$ be a window.
Then
\[ \lim_{R\to\infty} \frac{1}{\Vol (C_R)} \card (\vL(W) \cap C_R)
= \theta_H(W).\] \qed
\end{theorem}

Let $m$ be any positive integer and let $\bm:=\{1, \dots, m\}$. We assume that
we are given {\em disjoint} windows $W_1, \dots, W_m$, and then the corresponding
model sets
\[ \Lambda_j = \vL((W_j) \subset \RR^d \, .\]

Let $\gL := \bigcup_{i=1}^m \gL_i $ (a disjoint union). We assume given a set
$\bw = (w_1, \dots, w_m)$ of weights $w_j \in \RR$. For $x\in \RR^d$ we define
\[w(x) = \begin{cases}
w_j &\text{ if $ x \in \gL_j$} \\
0 &\text{otherwise.}
\end{cases}
\]
In our examples below we shall only use non-negative weights.

We call $(\gL_1, \dots \gL_m)$ with the weights $\bw$ a {\em weighted model set}.
More generally one would allow arbitrary translations of the colour component
sets $\gL_i$ provided the translated point sets do not overlap, but we have no need of this here. For notational
simplicity we shall use the symbol $\gL^{\bw}$ (or often simply $\gL$ is the context is clear) to  denote the coloured/weighted model set that we have just described.

The $n+1$-point {\bf correlation} ($n = 1,2, \dots$) of a model set
$\gL^{\bw}$ (or more generally any weighted locally finite subset of $\RR^d$) is the measure
on $(\Rd)^n$ defined by
\begin{eqnarray*}
\gamma_\gL^{(n+1)}(f)&=&\lim_{R\to\infty} \frac{1}{\Vol (C_R)}\sum_{y_1,\dots, y_n,x
\, \in \,C_R \cap \gL} w(x)w(y_1) \dots w(y_n) f(-x+ y_1, \dots,-x+ y_n)\\
&=&\lim_{R\to\infty} \frac{1}{\Vol (C_R)}\sum_{\stackrel{x\in C_R \cap \gL}{ y_1,\dots y_n \in \gL}} w(x)w(y_1) \dots w(y_n) f(-x+ y_1, \dots,-x+ y_n) \, ,
\end{eqnarray*}
for all continuous compactly supported functions $f$ on $(\RR^d)^n$, \cite{DM}.
The simpler second sum is a result of the compactness of the support of $K$ of
$f$ and the van Hove property of the averaging sequence $\{C_R\}$ (namely that
the measures of the $K$-boundaries of the sets $C_R$ have vanishing relevance relative to the total volume of $C_R$ as $R\to\infty$ .

Because model sets are Meyer sets, the sets of elements
$y_j-x $ which make up the values of the arguments of $f$ occuring
in the sums lie in the {\em uniformly discrete} set $\gL - \gL$.
For any $\bz = (z_1, \dots, z_n) \in (\gL -\gL)^n$ and any
$\bk = (k(0), \dots, k(n)) \in \bm^{n+1}$ we can count the occurrences of
$z_1, \dots, z_n$ in the form $-x +y_1, \dots, -x +y_n$ where
$ x \in \gL_{k(0)}$, $y_j \in \gL_{k(j)}$, $j = 1, \dots, m$, namely
\begin{eqnarray*}
\lefteqn{\frq_{\bk}(\{z_1, \dots, z_n\}) =} \\
 && \lim_{R\to\infty} \frac{1}{\Vol(C_R)} \card \{x \in C_R\, : \, x \in \gL_{k(0)}, y_j \in \gL_{k(j)}, x = -z_j +y_j,j=1, \dots, m \}\\
&&=\mbox{  } \lim_{R\to\infty} \frac{1}{\Vol(C_R)} \card \{x \in L\cap C_R\, : \, x^\star \in
W_{k(0)} \cap \bigcap_{j=1}^m (-z_j^\star + W_{k(j)})\}\\
&&= \theta_H\left(W_{k(0)} \cap \bigcap_{j=1}^m (-z_j^\star + W_{k(j)})\right)
 \, ,
\end{eqnarray*}
where we have used the uniform distribution theorem.

Hence for model
sets we find that all these correlation measures exist and
\begin{equation}\label{eqcorrelation3}
  \gamma_\gL^{(n+1)} = \sum_{z_1,\dots,z_n\in \gL -\gL}
 \left( \sum_{\bk\in\bm^{n+1}} w_{k(0)} \dots w_{k(n)} \, \frq_{\bk}(z_1, \dots, z_n) \right)
 \, \delta_{(z_1,\dots,z_n)} \,.
\end{equation}

\section{A periodic example} \label{periodic}
Let $N$ be a positive integer and let $H= \ZZ/N\ZZ$. Then we have the
cut and project scheme:

\begin{equation} \label{cpFiniteScheme}
   \begin{array}{ccccc}
      \RR & \stackrel{\pi_{1}}{\longleftarrow} &
        \RR \times \ZZ/N\ZZ & \stackrel{\pi_{2}}
      {\longrightarrow} & \ZZ/N\ZZ   \\
      &&  \cup \\
      \ZZ&\stackrel{\simeq}{\longleftrightarrow} & \widetilde \ZZ &\qquad &\qquad\\
      x & \longleftrightarrow& (x,x_N) & \mapsto & x_N \,,
   \end{array}
\end{equation}
where $x_N:= x \mod N$ and $\widetilde \ZZ := \{(x,x_N) \,:\, x \in \ZZ\}$.
Let $\bm$ be as above. We assume that we are given disjoint subsets (windows) $A_1, \dots, A_m \subset \ZZ/N\ZZ$ and corresponding colour weights $w_1, \dots, w_m$. The
corresponding model set is $\gL = \cup \gL_j$ where $\gL_j = \vL(A_j) = \{ x \in \ZZ\,:
\, x_N \in A_j \}$, $j=1, \dots m$. Each colour set $\gL_j$ is periodic, repeating modulo $N$, containing all those integers congruent$\mod N$ to an element of $A_j$.

For each
$j= 0, \dots N-1$ define
\begin{equation} \label{cDef}
c_j = \begin{cases}
w_k & \text{if $j \in A_k$}\\
0 &  \text{otherwise.}
\end{cases}
\end{equation}
A given pattern $(r_1, \dots r_n)$ of integers modulo $N$ must occur in $\gL$ in the form
\[s + r_1, \dots , s+ r_n \;\mod N, \quad s=0,1, \dots, N-1  \, .\]
If it occurs with colours $k(1), \dots, k(n)$  and $k(0)$ is the colour
of $s$ then $r_j \in \gL_{k(j)}  - \gL_{k(0)}$ for all $j$, and the weighting is
\[w_{k(0)} \dots w_{k(r)} = c_s c_{s+r_1} \dots c_{s+r_n} \,.\]
Thus, for this weighted model set we have
\[ \gamma^{(n+1)}_\gL = \frac{1}{N}\sum_{(r_1, \dots, r_n)} M_n(r_1, \dots, r_n) \delta_{(r_1, \dots,r_n)} \]
where the sum runs over all possible patterns of length $n$ and where
\begin{equation} \label{defM}
M_n(r_1, \dots, r_n) := \sum_{s=0}^{N-1} c_s c_{s+r_1} \dots c_{s+r_n} \,.
\end{equation}

By the above arguments,

\begin{equation}\label{Mformula}
M_n(r_1, \dots, r_n) = \sum_{\bk\in \bm^{n+1}} w_{k(0)} \dots w_{k(n)} \, \theta_{\ZZ/N\ZZ}\left(A_{k(0)} \cap \bigcap_{j=1}^n (-r_j^\star + A_{k(j)})\right) \,.
\end{equation}

\section{Aperiodic Examples} \label{ae}
Start with the cut and project scheme $\CalS$ of \S\ref{ms}, along with the corresponding notation. Let
\[\alpha : L \longrightarrow \ZZ/N\ZZ \]
be any surjective homomorphism with the property that
\[\{ (x^\star, \alpha(x)) : x \in L\} \; \text{is dense in $H\times \ZZ/N\ZZ$} \,.\]
Form the new cut and project scheme $\CalS^e = (\RR^d, H \times \ZZ/N\ZZ,
\cL^e)$
where $\cL^e := \{(\tilde x, \alpha(x)) : x \in L\}$:
\begin{equation} \label{ExtcpScheme}
   \begin{array}{ccccc}
      \RR^d & \longleftarrow &
        \RR^d \times H \times \ZZ/N\ZZ &
     \longrightarrow & H \times \ZZ/N\ZZ   \\
      &&  \cup \\
      L&\stackrel{\simeq}{\longleftrightarrow} & \cL^e &\qquad &\qquad \\
      x& \longleftrightarrow & (\tilde x, \alpha(x))  & \mapsto & (x^\star, \alpha(x)) \quad.
   \end{array}
\end{equation}
$\cL^e$ is clearly discrete and since the index $[\cL^e : (\{\tilde x \in \cL : \alpha(x) =0\},0)]$ is finite and
the group $\ZZ/N\ZZ$ is finite, the quotient of $\RR^d \times H \times \ZZ/N\ZZ$
by $\cL^e$ is compact. In short, $\cL^e$ is a lattice in $\RR^d \times H \times \ZZ/N\ZZ$.

Now we let $\bm$ and the sets $A_j$ be as in \S\ref{periodic}.
Let $W$ be a non-empty window in $H$ and set $W_j:= W \times A_j \subset
H \times \ZZ/N\ZZ$.  This produces from $\CalS^e$ coloured model sets
\begin{equation} \label{extModelSet}
\gL^e_j = \vL^e(W_j) \,.
\end{equation}
 Let $\gL^e := \cup_{j=1}^m \gL_j^e$. Notice that the
actual points of the model sets involved here form a subset of the model set determined by the original cut and project
scheme $\CalS$, whereas the colours are being determined by the periodic
scheme.

The $(n+1)$-point correlation for $\gL^e$ is
\begin{equation}\label{eqcorrelation4}
  \gamma_\gL^{(n+1)} = \sum_{z_1,\dots,z_n\in \gL^e -\gL^e}
 \left( \sum_{\bk\in\bm^{n+1}} w(k(0)) \dots w(k(n)) \, \frq_{\bk}(z_1, \dots, z_n) \right)
 \, \delta_{(z_1,\dots,z_n)}
\end{equation}
where
\[
\frq_{\bk}(\{z_1, \dots, z_n\}) = \theta_{H\times\ZZ/N\ZZ} \left(W_{k(0)} \cap \bigcap_{j=1}^n (-z_j^\star + W_{k(j)})\right)
 \, .\]

 Let $z_j^\star =(v_j^\star,r_j^\star) \in H \times \ZZ/N\ZZ$. Then for $(q,r) \in H \times \ZZ/N\ZZ$,
 \begin{eqnarray*}
 (q,r) \in W_{k(0)} &\cap& \bigcap_{j=1}^n (-z_j^\star + W_{k(j)})\\
 &\Leftrightarrow& (q,r) \in W \times A_{k(0)} \;\text{and} \; (q,r) \in (-v_j^\star, -r_j^\star)
 + W\times A_{k(j)} \\
 &\Leftrightarrow& q \in W\cap \bigcap(-v_j^\star +W) \; \text{and} \;
 r \in A_{k(0)} \cap \bigcap(-r_j^\star  + A_{k(j)}) \,.
 \end{eqnarray*}
 Thus the (relative) frequencies are given by
 \[ \frq_{\bk} (z_1, \dots, z_n) =
 \theta_H \left(W \cap \bigcap_{j=1}^n (-v_j^\star + W)\right)
 \theta_{\ZZ/N\ZZ} \left(A_{k(0)} \cap \bigcap_{j=1}^n (-r_j^\star +A_{k(j)})\right)\,.
 \, \]
 The first term of this factorization is independent of $\bk$ and as a consequence we can rewrite \eqref{eqcorrelation4} as
 \begin{equation}\label{eqcorrelation4}
  \gamma_\gL^{(n+1)} = \sum_{z_1,\dots,z_n\in \gL^e -\gL^e}
 \theta_H \left(W \cap \bigcap_{j=1}^n (-v_j^\star + W)\right)M_n(r_1, \dots, r_n)
 \, \delta_{(z_1,\dots,z_n)}
\end{equation}

It is not particularly important for our purposes that the frequencies here be absolute. As we already pointed out, that depends on normalizing the Haar measures so that the corresponding Haar measure on $(\RR^d \times H \times \ZZ/N\ZZ)/ \cL^e$ has total measure equal to $1$. What is important is to realize that if we colour and weight by a second set of weights
$w_1', \dots, w_m'$ and a second set of windows $A'_1, \dots, A'_m$
so that the expressions of \eqref{Mformula} are equal for some $n$, then also
the $(n+1)$-point
correlations of the two corresponding weighted model sets $\gL^e$ arising from
using one or the other of these two colour/weighting schemes will be the same.
This information is contained in \eqref{defM}, and it is this form that we shall
see in the examples.

It is interesting to note here that the formula for the frequencies makes it look as if we are dealing with
a simple product structure. However, the points $z_j =(v_j, r_j)$ are not truly from an unrestricted product. In fact,  $r_j = \alpha(v_j)$. The reason for the frequencies to be
given as they are is that the lattice $\cL^e$ already has this special structure
built into it. We have assumed that its image is dense in
$H\times \ZZ/N\ZZ$, so we have a cut and project scheme, and this allows us to use the uniform distribution of model sets to derive the frequencies in terms of the measures of the windows.

\section{Examples}\label{CE}

With $N = 6$,  there are two sets of weights
\begin{eqnarray} \label{weights}
\ws_1 &:=& [11,25,42,45,31,14] \nonumber\\
\ws_2 &:=& [10,21,39,46, 35, 17]
\end{eqnarray} which, when used to weight the sets
$A_j = A'_j = \{j\} \mod 6$, $j=0, \dots, 5$, determine identical results on the left
side of \eqref{defM} for $n=1,2,3,4$, see  \cite{GM} \S5.3. It follows from
our discussions that the corresponding weighted model sets built
in \eqref{extModelSet}, though quite different, nonetheless have equal $2,3,4,5$-point
correlations.

Although the information needed to show
that the sums arising in \eqref{defM} from these two sets of weights are the same
is implicit in
\cite{GM}, and although it would be easy to check the result on a computer,
it is interesting to see what lies behind this.

Given a collection of disjoint subsets $A_1,\dots, A_m$ of $\ZZ/N\ZZ$ and a set of
weights $w_m$, $k = 1, \dots m$, we define
$\bc= (c_0, \dots, c_{N-1})$, where the $c_j \in \ZZ$,  using \eqref{cDef}. The corresponding weighted Dirac comb is
\[ D := \sum_{j=0}^{N-1} c_j \delta_j \, , \]
and its Fourier transform is $\widehat D$ given by
\[ \widehat D(k) = \sum_{j=0}^{N-1} c_j \exp\left({\frac{-2 \pi i jk}{N}}\right) = P(w^k) \]
where $w := e^{-2 \pi i/N}$ and $P = P^{\bc}$ is defined by
\[P(x) = \sum_{j=0}^{N-1} c_j x^j \,.\]

The pattern frequency of $(l_1, \dots, l_n)$ in the weighted periodic point set determined by our choice of the cut and project scheme \eqref{cpFiniteScheme}, the windows
$A_1, \dots A_m$, and the weighting system $\bc$ is,
up to the appropriate normalization factor,
given by \eqref{defM}:
\begin{equation}\label{Mfunction}
M_n(l_1, \dots l_n) = \sum_{l=0}^n c_l c_{l+l_1} \dots c_{l+ l_n} \,.
\end{equation}
In this way we have a function
\[M_n : (\ZZ/N\ZZ)^n  \longrightarrow \RR \, .\]
A straightforward calculation of the Fourier transform of $M$ leads to
\begin{eqnarray}
\widehat {M_n}(k_1, \dots, k_n) &=& \widehat{D} (k_1) \dots \widehat{D} (k_n)
\widehat{D} (-(k_1 + \cdots + k_n))\nonumber \\
&=& P(w^{k_1}) \dots P(w^{k_n}) P(w^{-(k_1 +\cdots + k_n)}) \,.
\end{eqnarray}

Since $M_n$ and $\widehat{M_n}$ deterimine each other, knowing one is the same
as knowing the other. In particular, if two weighting systems determine
$\bc$ and $\bc'$, and these determine the same functions
$ P(w^{k_1}) \dots P(w^{k_n}) P(w^{-(k_1 +\cdots + k_n)})$ for some $n$,
then they also produce the same $n+1$-point pattern frequencies.

Let us apply this analysis to the two weighting systems for $\ZZ/6\ZZ$ given
in \eqref{weights}. Here the two corresponding polynomials are
\[ P_1(x) :=(x+1)(x^2+x+1)(2x^2+5)(3x+1) =: Q_1(x)(3x+1) \]
and
\[P_2(x) :=(x+1)(x^2+x+1)(2x^2+x +4)(3x+1) =: Q_2(x)(3x+1) \, ,\]
Since we are only interested in the values of these at powers of $w = e^{-2\pi i/6}$
we may alter $P_1$ and $P_2$ so that their exponents of $x$ are reduced modulo $6$,
so that they all lie in the range $0, \dots, 5$. Having done this one checks that these indeed
produce the coefficients $c_0, \dots, c_5$ given by $\ws_1, \ws_2$.
We also observe that by the construction these polynomials vanish at
$w^2, w^3, w^4$. Also the two polynomials are equal for $x=1$ and furthermore,
$w^4 Q_1(w^{-1}) =  Q_2(w)$ and $w^4 Q_2(w^{-1}) = Q_1(w)$. This  means
that $Q_1(w)P_1(w^{-1}) = Q_2(w)P_2(w^{-1})$ and hence $P_1(w)P_1(w^{-1}) = P_2(w)P_2(w^{-1})$.

Now consider the situation of the $5$-point correlations. We have to look at
\begin{equation}
P(w^j)P(w^k)P(w^l)P(w^r) P(w^{-(j+k+l+r)})
\end{equation}
for the two polynomials and show that these values are the same for all possible values of $j,k,l,r$. Assuming that none
of these indices is $0$ mod $6$ and taking into account the vanishing properties above and the symmetry of the four indices $j,k,l,r$, we have the following table of possibilities:

\begin{center}
\begin{tabular}{cccc|c}
j & k & l & r & -(j+k+l+r)\\ \hline
1 &1 & 1 & 1 & -4 \\
1 &1 & 1 & -1 & -2\\
1 &1 & -1 & -1 & 0\\
1 &-1 & -1 & -1 & 2\\
-1 &-1 & -1 & -1 & 4\\
\end{tabular}
\end{center}

Given the vanishing properties of $P_1,P_2$ , only the middle case is
non-trivial. However, there the facts that $1$ and $-1$ occur equally often
and that the polynomials are equal at $w=1$ gives the result.

If one of the indices, say $r$ is zero, then we are in the situation of
the $4$-point correlation. There the only non-obvious case is when
$i,j,k = \pm1$ and all three are not equal. Then along with $-(i+j+k)$ we have
$1$ twice and $-1$ twice  and so again the products are equal.

The situation for the $2$- and $3$-point correlations is equally simple.

We thus have:

 \begin{theorem} \rm{Given the cut and project scheme \eqref{cpScheme} and a nonempty window
 $W \subset H$, then the two (different!) aperiodic weighted model sets arising from
 the cut and project scheme \eqref{ExtcpScheme} and the two mod 6 weighted colourings given by $\ws_1$ and
 $\ws_2$ \eqref{weights} have the same $2,3,4,5$-point correlations. }\qed
 \end{theorem}

\subsection{A 2-dimensional example}

\begin{figure}
\centering
\includegraphics[width=12cm]{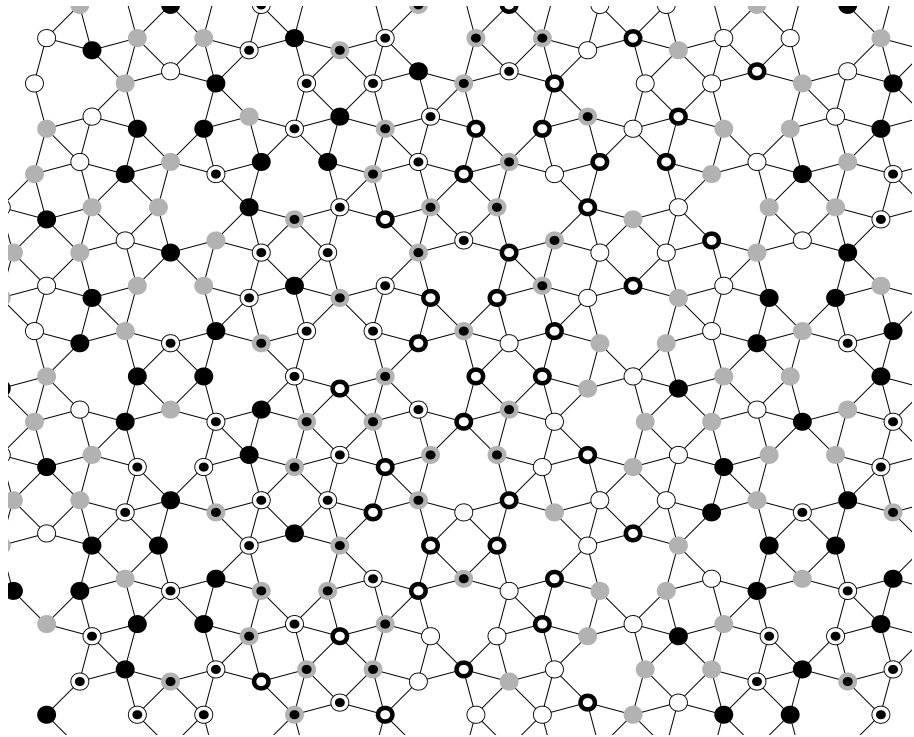}
\caption{A fragment of the shield tiling with a $6$-colouring}
\label{tshield}
\end{figure}

The STS tiling, or shield tiling, is an aperiodic substitution tiling discovered by
F. ~G\"{a}hler \cite{GF88}. It consists of three types of tiles: squares, triangles and asymmetric hexagons looking like shields, see Fig.~\ref{tshield}. The vertices of an STS tiling can
be realized as a model set $\CalS = (\RR^2 \times \RR^2, \cL)$ whose lattice
can be described as
\[\cL =  \{(k_1,k_2,k_3,k_4) \,: \, 2 k_i \in \ZZ \quad \mbox{for all $i$, all of them even or all of them odd.} \} \, .\]

This lattice has a automorphisms $C$ of order $12$. In fact, Lie theorists will recognize the lattice as the root lattice of type $F_4$ and $C$ can be chosen as any one
of its (conjugate) Coxeter transformations \cite{Bourbaki}. The four eigenvectors of $C$ lead to two real $C$-invariant spaces
and it is the projections onto these that create the cut and project scheme. In each of these $C$ appears as a rotation of order $12$. The
window is a regular dodecagon $W$, displaced generically to avoid the projections
of any of the lattice points of $\cL$ falling on the boundary of $W$ \cite{GF88}.

Inside $\cL$ we have the sublattice of index $2$ obtained by restricting the
vectors to have integral components (the $D_4$ lattice). Inside that there is a sublattice
$\cL_0$ consisting of those vectors the sum of whose components is congruent
to $0$ modulo $3$. Then $\cL/\cL_0 \simeq \ZZ/6\ZZ$ provides
us with a homomorphism $\alpha : \cL \longrightarrow \ZZ/6\ZZ$ with which
we can carry out the construction of \S\ref{ae}.

The resulting colouring on the shield tiling is indicated in the Fig.~1 with the
different colours indicated by different symbols. Symbols that differ only by the presence or
absence of a centre dot correspond to  colours differing by $3$ modulo $6$. There are no $6$-colourings that respect the rotation $C$. Instead, what one can see here is the bands of like-colours moving in roughly a north-by-northeast direction (the shortest edge vector in this direction has degree $0$)
and the cycling through the six bands as one moves in the normal directions.
According to the theory above, each of the two weighting systems of
\eqref{weights} can be applied
to this model set and the resulting sets will be indistinguishable from the point of view of their $2,3,4,5$-point correlations. The
results of \cite{GM} imply that they {\it are} distinguishable by their $6$-point correlations.

The results of \cite{GM}, on which the construction of this paper depends, seem
not to have been generalized to dimensions greater than one. It would be interesting to
do this since it would probably give rise to other even more interesting examples of distinct model
sets with many identical correlations.

\section{A stochastic interpretation}
We finish by pointing out that it is possible to place the results and examples described in this paper into a stochastic setting.
We use the same ingredients as before:
an unweighted model set $\gL$, a homomorphism
$\alpha$, and a set of weights. We assume now that all the weights are non-negative
and scaled so that they all lie in the range
$[0,1]$. We imagine now that the points of the basic unweighted model set $\gL$
are selected or not selected on the basis of independent random choices
at each site, the probability of being selected being $w_j$ if the point is of colour $j$.
The resulting structure is a point process, each event being the outcome of
independent Bernoulli trials made at every one of the sites according to the probabilities
given by the weights. The moment measures that we have described then
describe the expected values of the patterns can occur in the point process. The
consequences of this for the two-point correlation and the corresponding diffraction measure can be explicitly determined from \cite{BM},Thm.~2. In particular the diffraction consists, almost surely, of a pure point part plus a continuous constant background.

\section{Acknowledgments}
RVM acknowledges the support of the Natural and Engineering Research
Council of Canada.
X.D. would like to thank the CRC 701 for
support during a four month visit to Bielefeld.

\end{document}